# A 2.7 µm Backward Wave Optical Parametric Oscillator Source for CO$_2$ Spectroscopy

ADRIAN VÅGBERG[1,†,*], MARTIN BRUNZELL[1,†], MAX WIDARSSON[2], PATRICK MUTTER[1,2], ANDRIUS ZUKAUSKAS[1], FREDRIK LAURELL[1], AND VALDAS PASISKEVICIUS[1]

[1]*Department of Applied Physics, KTH Royal Institute of Technology, Roslagstullsbacken 21, 106 91 Stockholm, Sweden*
[2]*SLF Svenska Laserfabriken AB, Ruddammsvägen 49, 114 19 Stockholm, Sweden*
[†]*The authors contributed equally to this work.*
*\*avag@kth.se*

**In this study, we demonstrated the direct use of an inherently narrowband backward wave in the mid-infrared for CO$_2$ and H$_2$O vapor spectroscopy in ambient air. This wave is generated in a backward wave optical parametric oscillator (BWOPO) pumped by a multi longitudinal mode laser at 1030 nm, eliminating the need for additional spectral narrowing or wavelength stabilization. A full characterization of the source is presented, revealing a central output at 2712 nm, showcasing temperature tuning of -1.77 GHz/K, and achieving an output pulse energy of 2.3 µJ. Novel methods are introduced for measuring the linewidth and wavelength stability using lab air. These approaches demonstrate a narrow output of 43 pm and establish an upper limit of stability at 65 MHz, with no active means of stabilization. These findings underscore the potential of BWOPOs as a robust platform for future DIAL systems.**

Carbon dioxide (CO$_2$) stands as one of the foremost greenhouse gases, exerting a profound impact on global climate dynamics and environmental equilibrium. The need to monitor atmospheric CO$_2$ concentrations precisely and continuously has never been more pressing [1]. Optical systems offer a promising avenue for remote sensing, leveraging technologies like differential absorption lidar (DIAL) to achieve high spatial resolution [2]. All DIAL setups rely on a complex laser source able to produce a spectrally narrow, pulsed output of two closely spaced wavelengths – often denoted as the *on* and *off* pulses. The *on* pulse is placed where absorption is high and *off* where it is low. By comparing the difference in intensity of the backscattered light and time gating the signal, a spatially resolved concentration can be extracted [3–5]. The main challenge high-resolution DIAL systems face is the absence of adequate sources and receivers at the target wavelength [6]. CO$_2$ exhibits many rotational vibrational transitions which can be used in gas spectroscopy [7]. In situ monitoring can only be done in a real atmospheric setting, implying interfering absorption lines due to the presence of several gases with partially overlapping lines. In practice, water vapor is a significant contributor to interference in the IR region because of its strong absorption [8]. To avoid this, and to increase absorption length, most atmospheric DIAL systems targeting CO$_2$ operate in a wavelength range around 2.05 µm. Here the absorption cross-section of water vapor for the *on* and *off* lines can be matched and remains orders of magnitude lower than for CO$_2$ [9,10].

A significant challenge in remote sensing arises from the lack of overlap between strong rotational vibrational lines and laser transitions. One approach is to use solid state lasers, and specifically, Ho:YLF or Tm:Ho:YLF to target the 2.05 µm transition [11,12]. Other wavelengths for long-distance CO$_2$ DIAL have been targeted using difference frequency mixing [13] and optical parametric oscillators (OPOs) employing pumping with commercial injection-seeded lasers at 1 µm [14-16]. While traditional OPO systems have demonstrated their ability to deliver high-energy outputs and wide tunability, they suffer from bulkiness and sensitivity to vibrations. That poses significant challenges, especially in applications such as satellite-based remote sensing [17].

Advancements in quasi-phase matched (QPM) materials structuring technology have enabled the development of backward wave optical parametric oscillators (BWOPOs) using periodically poled potassium titanyl phosphate (PPKTP) [18] with sub-micron poling periods. BWOPOs do not need an optical cavity and have the inherent property of narrowband backward wave generation even for a multimode pump, as well as high efficiency and high temperature stability [19, 20]. These properties make BWOPO suitable for compact and environmentally robust sources of radiation.

In this study, we explore the advantages of employing BWOPOs for DIAL. This is the first such demonstration where we use ambient CO$_2$ and H$_2$O vapor with partially overlapping spectral lines. Moreover, in this work we intentionally use a multi-longitudinal mode pump laser at 1.03 µm that would render the other OPO based techniques unsuitable.

While the absorption of CO$_2$ at 2.05 µm is well-suited for long-distance atmospheric measurements, it provides very limited absorption in a laboratory environment where sufficiently long path lengths are unattainable. Consequently, this study targets the strong absorption in the 2.7 μm band [7] in order to investigate the applicability of BWOPOs for DIAL. BWOPOs allow for the design of QPM periods to generate wavelengths across the entire transparency window, including 2.05 μm. Eqs. (1,2) illustrate the tuning of the backward and forward waves relative to the pump [18]:

$$\frac{\partial \omega_f}{\partial \omega_p} = \frac{v_{g,f}(v_{g,p} + v_{g,b})}{v_{g,p}(v_{g,b} + v_{g,f})} = 1 + \epsilon \quad (1)$$

$$\frac{\partial \omega_b}{\partial \omega_p} = \frac{v_{g,b}(v_{g,p} - v_{g,f})}{v_{g,p}(v_{g,b} + v_{g,f})} = -\epsilon \quad (2)$$

Here $\omega$ is the angular frequency of the light and $v_{g,p}$, $v_{g,f}$, and $v_{g,b}$ are the group velocities of the pump, forward and backward waves. The value of $\epsilon$ is usually on the order of $10^{-3}$ and in this work it is calculated to be $4.4 \cdot 10^{-3}$ using the Sellmeier's equation for KTP from [21]. This also implies that the backward wave linewidth can be orders of magnitude smaller than the pump. For nanosecond pulses the spectral width will ideally be limited by the Fourier transform of the temporal trace. A corollary to this is the inherent wavelength stability of the backward wave, where fluctuations in pump wavelength will be suppressed by the factor of $\epsilon$. Ensuring wavelength stability is critically important for achieving high accuracy in determination of concentrations in gas spectroscopy. Typically, obtaining the required precision necessitates intricate and costly frequency locking mechanisms [17]. In this context, the inherent wavelength stability of the BWOPO significantly reduces the setup complexity.

Fig. 1. shows the experimental setup, where the pump laser is guided into a BWOPO, realized with a PPKTP crystal with sub-micron period (provided by SLF Svenska Laserfabriken AB).

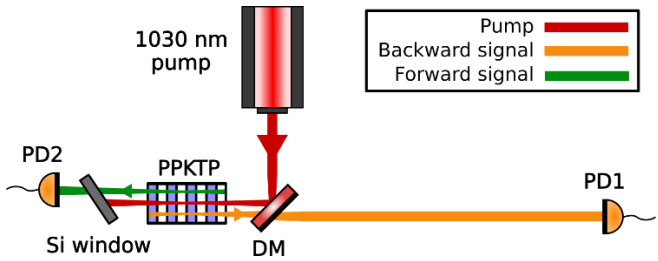

Fig. 1. Schematic of the experimental setup for the backward wave $CO_2$ absorption measurements.

The pump laser (Thorlabs QSL103A), emitting at 1030 nm, had a repetition rate of 9 kHz, pulse length of 380 ps, and a multi-longitudinal mode spectrum with FWHM width of 360 GHz. The PPKTP crystal was mounted in a temperature-controlled crystal mount. In the crystal, the pump is converted into a copropagating forward wave centered around 1661 nm and a counterpropagating backward wave at 2712 nm. The backward wave was extracted through a dichroic mirror, where it could be characterized and used for spectroscopy.

The measurements were performed by sweeping the temperature of the crystal from 15 °C to 90 °C in steps of 0.1°, resulting in an operational spectral span between approximately 2711.4 nm and 2714.6 nm. The beam path from the PPKTP crystal to the power meter PD1 (Thorlabs S401C) was approximately 290 cm. A Ge photodiode PD2 (Thorlabs S122C) was used to measure the power of the unattenuated forward signal. The Si window acted as a long-pass filter, removing unwanted residual pump light.

The expected transmittance in ambient air was first simulated using the HITRAN Python API [22]. The simulated absorption cross-sections for various atmospheric gases in air are summarized in Fig. 2 (a).

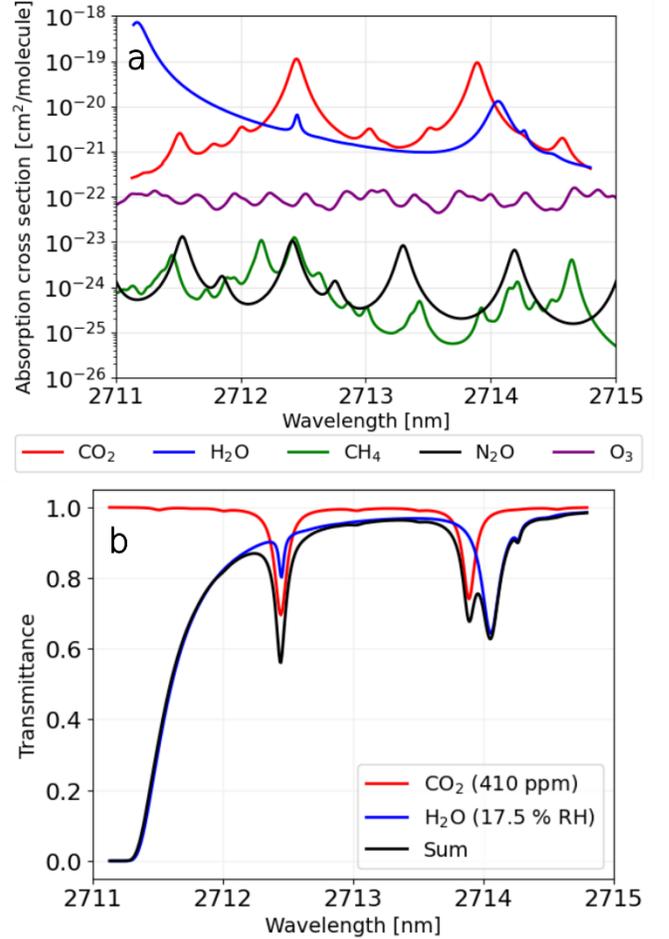

Fig. 2. (a) Simulated absorption cross-section of relevant atmospheric gases in air and (b) simulated transmittance spectrum for a 290 cm beam path in the laboratory.

From this, it is apparent that we can limit our simulations to $CO_2$ and water vapor since the other trace gases will have a negligible impact on the absorption. The transmitted power of both the forward and backward wave were measured, as indicated in Fig. 1. For every generated forward photon, we have necessarily also generated a backward photon. Thus, we can use the ratio of the measured powers to represent the transmittance. The benefit of this approach is that the impact of power fluctuations in the pump can be largely compensated for. The pulse energies of the backward wave exceeded 2.3 μJ in this experiment. The measured transmission is depicted in Fig. 3 along with a matched simulation. In order to compare the collected data with the simulation, the data has been normalized so that the maximum transmissions coincide. The similarity to the simulated spectrum is obvious besides the weak oscillations, and slight decrease in power, at wavelength longer than 2714 nm, i.e. the highest temperatures for the BWOPO crystal. We suspect that these oscillations arise from the detection scheme utilized, wherein the 2.7 μm detector exhibits a significantly longer rise time (~ 1 s) compared to the forward wave detector (< 1 μs). Our pump laser exhibits fast power fluctuations, and the comparably slow thermal detector effectively functions as a low-pass filter, distorting the collected data relative to the much faster forward beam detector.

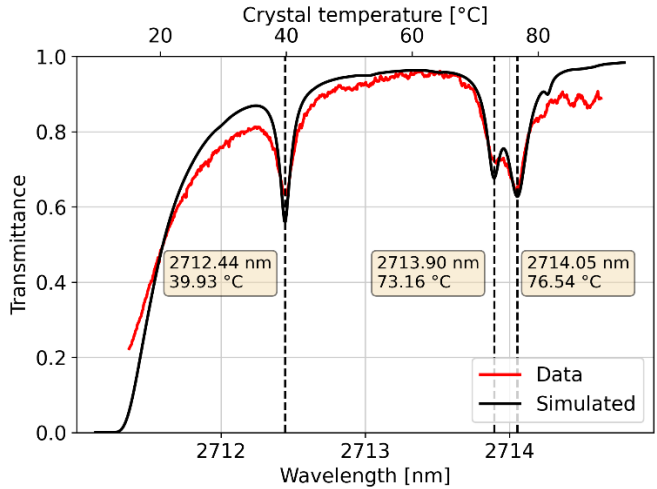

Fig. 3. Measured transmittance overlaid on the predicted spectrum, with the temperature and wavelength of three distinctive peaks marked.

When the measured data is overlaid on the predicted transmittance spectrum, we can use the temperature and wavelength of the three distinctive peaks to determine the tuning parameter of the BWOPO to be -1.79 GHz/K. This is very close to the predicted -1.77 GHz/K we get when using the refractive index data for KTP given by Katz *et al.* [21], the thermo-optic coefficients measured by Emanueli and Arie [23], and the thermal expansion coefficients determined by Pignatiello *et al.* [24]. The wavelength position of the absorption peaks is determined by leveraging the extensive tabulated absorption data [22]. The number of $CO_2$ peaks visible in the transmission spectrum were in our setup limited by the operation temperature of the crystal mount. To independently test the accuracy of the thermal tuning determination from the absorption peaks we used an optical spectrum analyzer (OSA) (Yokogawa AQ6376E), that resulted in a tuning rate of -1.75 GHz/K. The findings are summarized in Fig. 4 where a good agreement can be found for all measurements and simulations, reiterating the accuracy of the simple approach using ambient air absorption $CO_2$.

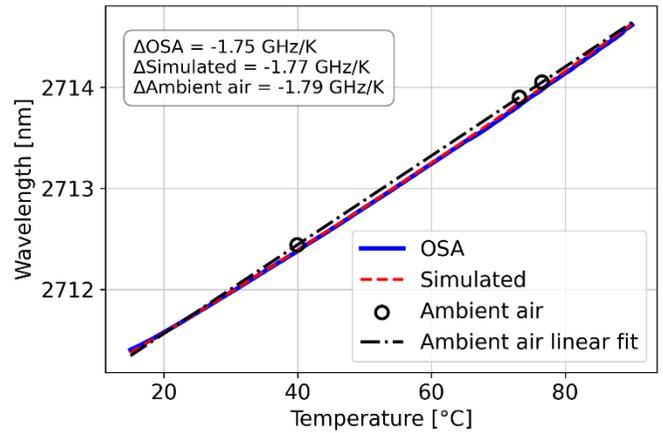

Fig. 4. Temperature tuning rate measurements compared to simulations. Two measurements were employed, one using an OSA and the other leveraging the known $CO_2$ and $H_2O$ spectral absorption lines.

The spectral bandwidth of the BWOPO backward wave was below resolution of the OSA (50 pm). Therefore, an alternative method for determining the linewidth by comparing the simulated and measured transmittance spectra was used. This was done by convolving the theoretical spectrum with Gaussian functions of different widths to emulate the linewidth of the source. A method of least squares was used to ascertain the best fit to the measured data, which corresponded to a full width at half maximum of 43 pm, or 1.75 GHz. The resulting convolution is overlaid with the measured data and displayed in Fig. 5. A rough estimate of the lower limit on the linewidth of the backward wave is the Fourier transform bandwidth of the pump pulse which is about 1.16 GHz. However, the BWOPO backward wave pulses, in general, are shorter than the pump pulses due to oscillation buildup time as in all parametric oscillators [19]. Therefore, the measured bandwidth of 1.75 GHz is likely close to transform-limited.

Given the limited resolution of the OSA, it was unsuitable for determining the spectral stability of the BWOPO backward wave. Instead, we employed the $CO_2$ lines directly. The method involved continuous measurement of the backward BWOPO intensity at two distinct parts of the transmission curve, followed by extraction of the relative errors (standard deviation divided by the mean). We chose spectral sections where the transmission dependence on wavelength was minimized (transmission *plateau* in Fig. 6) and maximized (transmission *wing* in Fig. 6). As shown in inset of Fig. 6 the wavelength of 2713.35 nm corresponded to the plateau, and 2713.84 nm to the wing segment. The difference in the relative error would then be in part attributable to the center wavelength shifts. The power transmission derivative with respect to wavelength, proportional to the signal path length, was maximized by leveraging the full length of the optical table, resulting in an 11.5 m beam path. In the absorption wing segment, a 1 pm (41 MHz) shift in central wavelength corresponded to a transmitted power difference of 0.95% (see Fig. 6). To mitigate power instability-related noise from the source, a reference power was measured in the BWOPO forward beam, facilitating pump power fluctuation-compensated signal acquisition.

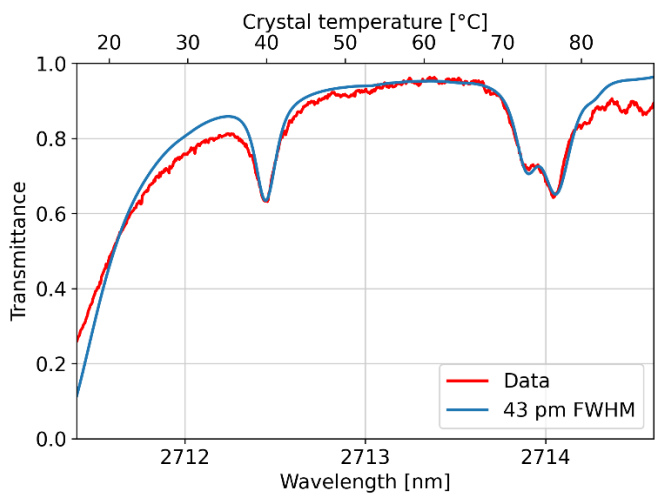

Fig. 5. Captured transmittance data compared to the predicted transmittance convolved with a 43 pm FWHM Gaussian.

Given the difference in rise time between the detectors for the forward and backward measurements, additional lowpass filtering was necessary post-data collection to establish a normalization function aiding in minimizing pump-related noise. A set of 18 measurements, each spanning 100 minutes at each wavelength, were conducted. The results, depicted in Fig. 6, illustrate the relative power fluctuation for both the plateau and wing segments.

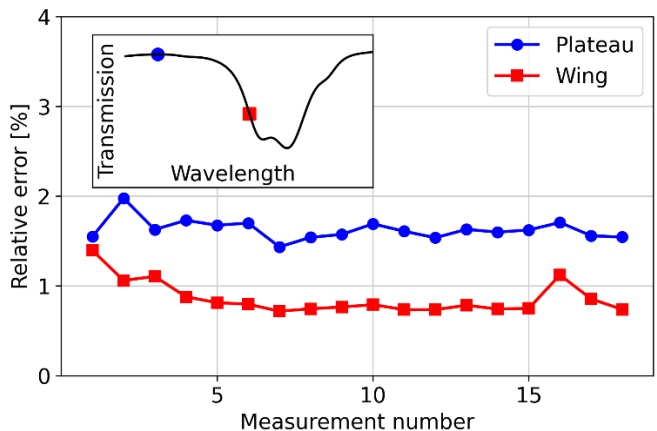

Fig. 6. Relative power error shown for the two regions where wavelength dependent transmission through air ($CO_2$ and $H_2O$ lines) are maximized and minimized, referred to as the wing and plateau. The placement of the wing and plateau is illustrated in the inset.

The analysis reveals comparable noise levels between the wing and the plateau, with the former even exhibiting slightly lower power fluctuations. This underscores the actual stability of the central wavelength. The measurement can only indicate the upper limit of the instability. By taking the highest fluctuation value at the wing of 1.5 % we get the wavelength fluctuation not exceeding 1.6 pm (65 MHz). The stability is most likely well below this value and achieved completely passively and with free-running multimode pump. Fluctuations stemming from pump variations can be mitigated by transitioning to a more stable, single-mode laser source.

In conclusion, we have demonstrated a nearly transform-limited and tunable pulsed laser source at 2712 nm that was realized using a multimode laser-pumped PPKTP BWOPO. This arrangement removes the need for a conventional OPO cavity and wavelength locking control and opens the possibility of creating a compact and robust source suitable for gas absorption spectroscopy and sensing. The results clearly show that very promising precision-tunable source for DIAL applications, including those on moving platforms can be realized using BWOPOs.

**Funding.** Strategiska innovationsprogrammet Smartare Elektroniksystem - en gemensam satsning av Vinnova, Formas och Energimyndigheten. Swedish Research Council grant 2023-04985.

**Disclosures.** The authors declare the following competing financial interest: SLF Svenska Laserfabriken AB fabricates and sells PPKTP which is used in this publication.

**Data availability.** Data underlying the results presented in this paper are not publicly available at this time but may be obtained from the authors upon reasonable request.